\documentclass[iop]{emulateapj}
\usepackage{rotating,epsfig,natbib}

\newcommand{\hrb}{HR8799~b}
\newcommand{\hrc}{HR8799~c}
\newcommand{\icarus}{Icarus}

\newcommand{\microns}{$\mu$m}
\newcommand{\chisq}{$\chi^2$}
\newcommand{\water}{H$_2$O}
\newcommand{\methane}{CH$_4$}
\newcommand{\co}{CO}

\newcommand{\teff}{$T_{\rm eff}$}
\newcommand{\lbol}{$L_{\rm bol}$}
\newcommand{\kzz}{$K_{\rm zz}$}

\newcommand{\logg}{$\log(g)$}

\newcommand{\mjup}{${\rm M}_{\rm Jup}$}

\makeatletter

\newcommand{\Rmnum}[1]{\expandafter\@slowromancap\romannumeral #1@}
\makeatother

\shorttitle{The Atmosphere of HR8799~b}
\shortauthors{Barman et al.}

\begin{document}
\bibliographystyle{apj}

\title{Simultaneous Detection of Water, Methane and Carbon Monoxide in the
Atmosphere of Exoplanet HR8799\MakeLowercase{b}}
 \author{Travis S. Barman}
 \affil{Lunar and Planetary Laboratory, University of Arizona, Tucson AZ 85721 USA \\
        Email: {\tt barman@lpl.arizona.edu}}

 \author{Quinn M. Konopacky}
 \affil{Center for Astrophysics and Space Science, University of California San Diego, La Jolla, CA, 92093, USA}
 \affil{Dunlap Institute for Astronomy and Astrophysics University of Toronto, Toronto, Ontario, Canada M5S 3H4}

 \author{Bruce Macintosh}
 \affil{Kavli Institute for Particle Astrophysics and Cosmology, Stanford University, Stanford, CA 94305 USA}
 \affil{Lawrence Livermore National Laboratory, 7000 East Avenue, Livermore, CA 94550, USA}

 \author{Christian Marois}
 \affil{NRC Herzberg Astronomy and Astrophysics, 5071 West Saanich Rd, Victoria, BC, Canada, V9E 2E7}

\keywords{planetary systems - stars: atmospheres - stars: low-mass, brown dwarfs}

\begin{abstract}
Absorption lines from water, methane and carbon monoxide are detected in the
atmosphere of exoplanet \hrb.  A medium-resolution spectrum presented here
shows well-resolved and easily identified spectral features from all three
molecules across the $K$ band.  The majority of the lines are produced by CO
and H$_2$O, but several lines clearly belong to CH$_4$.  Comparisons between
these data and atmosphere models covering a range of temperatures and gravities
yield $\log$ mole fractions of H$_2$O between -3.09 and -3.91, CO between -3.30
and -3.72 and CH$_4$ between -5.06 and -5.85. More precise mole fractions are
obtained for each temperature and gravity studied.  A reanalysis of $H$-band
data, previously obtained at similar spectral resolution, results in a nearly
identical water abundance as determined from the $K$-band spectrum.
The methane abundance is shown to be sensitive to vertical mixing and indicates
an eddy diffusion coefficient in the range of $10^6$ to $10^8$ cm$^2$ s$^{-1}$,
comparable to mixing in the deep troposphere of Jupiter.  The model comparisons
also indicate a C/O between $\sim$ 0.58 and 0.7, encompassing previous
estimates for a second planet in the same system, \hrc.  Super-stellar C/O could
indicate planet formation by core-accretion, however, the range of possible C/O
for these planets (and the star) is currently too large to comment strongly on
planet formation.  More precise values of the bulk properties (e.g., effective
temperature and surface gravity) are needed for improved abundance estimates.
\end{abstract}
\vspace{6pt}

\section{Introduction}

The HR8799 planetary system remains unique as the only system to have multiple,
directly imaged, planets \cite[]{Marois2008,Marois2010}. Four planets orbiting
HR8799 have been monitored regularly since their discovery,
providing the astrometric data needed to estimate their orbital properties and
masses.  The current astrometric data indicate that all four planets are less 
than 13 \mjup \cite[]{Marois2010, Fabrycky2010, Currie2011, Pueyo2014}.

The planets are frequently observed for the purposes of characterizing their
atmospheric properties and, in particular, their chemical compositions.  The
two outer most planets, b and c, have the most comprehensive wavelength
coverage including near-infrared (near-IR) spectroscopy at low, $R \sim 50 -
100$ \cite[]{Bowler2010, Barman2011a, Oppenheimer2013} and medium, $R \sim
4000$ \cite[]{Konopacky2013} resolutions.  The Gemini Planet Imager (GPI) has
measured low resolution spectra of planets c and d \cite[]{Macintosh2014,
Ingraham2014A}.  Spectroscopy of multiple planets orbiting HR8799 allows direct
comparisons of atmospheric compositions for a coeval set of planets formed from
the same protoplanetary disk.  The only other planetary system for which such a
comparison is currently feasible is our own Solar System.

\begin{figure*}[t]
\plotone{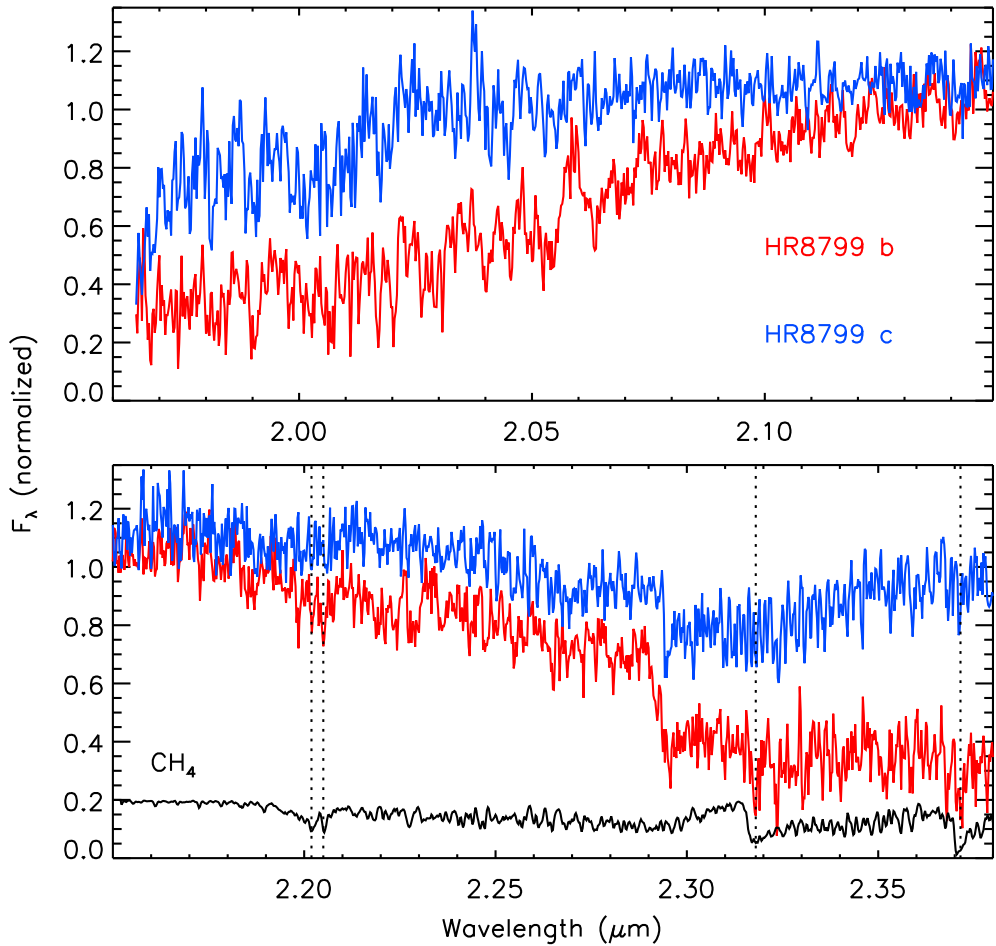}
\caption{
Top and bottom panels compare the medium-resolution near-infrared spectra of
\hrb\ (red, this work) and \hrc\ (blue; Konopacky et al. 2013), both
normalized to the same flux at 2.15 $\mu$m.  The lower panel includes a portion
of the methane absorption reduced to the same resolution and sampling as the
observed data and offset for comparison.  Regions of strong methane absorption
are indicated with vertical dotted lines.  \label{fig1}}
\end{figure*}

The giant planets in our Solar System were likely born with bulk compositions
determined by their initial location in the planetary disk and the specific
solid-to-gas (M$_{\rm solid}$/M$_{\rm gas}$) accretion history they
experienced. As these initially hot planets cooled with time, their atmospheres
experienced various levels of vertical mixing and condensation.  The
atmospheric composition may have been further altered by continued accretion of
solid bodies or mixing with a deep metal rich core.  Consequently, the
present-day mole fractions for important trace molecules (H$_2$O, CH$_4$,
CO$_2$, CO, NH$_3$, N$_2$) of giant planets in our Solar System are the result
of numerous chemical and physical processes.  Inferring elemental abundances
from molecular abundances requires a clear understanding of the atmospheric
chemical and dynamical history.  Even for Jupiter and Saturn only upper limits
on carbon-to-oxygen ratio (C/O) have been measured, a poignant reminder of how
challenging such measurements can be \cite[]{Wong2004,Visscher2005}.  Despite
this complex connection between present-day composition and the composition at
birth, it remains plausible that the present-day molecular abundances hold
clues pertaining to the formation history of our giant planets.  By observing
{\em young} giant planets, like those orbiting HR8799, billions of years of
atmospheric evolution that blurs the connection between atmospheric properties
and the formation process is avoided.

Presented below are new observations of \hrb\ that provide comparable
wavelength coverage, spectral resolution and signal-to-noise (SNR) as similar
observations of \hrc\ that revealed individual resolved water and carbon
monoxide absorption features \cite[hereafter K13]{Konopacky2013}.  For \hrb,
the mole fractions of these two molecules as well as methane are determined using
well-resolved spectral features.  From these mole fractions, the coefficient of
eddy diffusion (\kzz) and C/O are estimated.

\section{Observations and Spectrum Extraction}

\hrb\ was observed in 2013 on July 25, 26 and 27 (UT) with the OSIRIS
instrument \cite[]{Larkin2006} in the $K$ band using an identical instrument
configuration as previous work by this group \cite[hereafter B11]{Barman2011a}
resulting in 5.2 hrs of on-target integration time under good observing
conditions.  The data were calibrated and rectified to produce
three-dimensional basic calibrated data cubes \cite[]{Krabbe2004}.  The data
were initially binned to $R \sim 100$ and speckles suppressed following
B11\nocite{Barman2011a}.  As in B11\nocite{Barman2011a}, at low-$R$, the
spectrum SNR is limited by residual correlated speckle noise.  Despite the
increase in signal from combining the 2009 to 2013 observations, the updated low-$R$
spectrum has mostly unchanged uncertainties and matches the earlier
B11\nocite{Barman2011a} spectrum. The $K$-band spectrum was absolute flux
calibrated using the $K_s$ magnitude ($M_{\rm K_s} = 14.15 \pm 0.1$) reported in
B11\nocite{Barman2011a}.

\begin{figure}[t]
\plotone{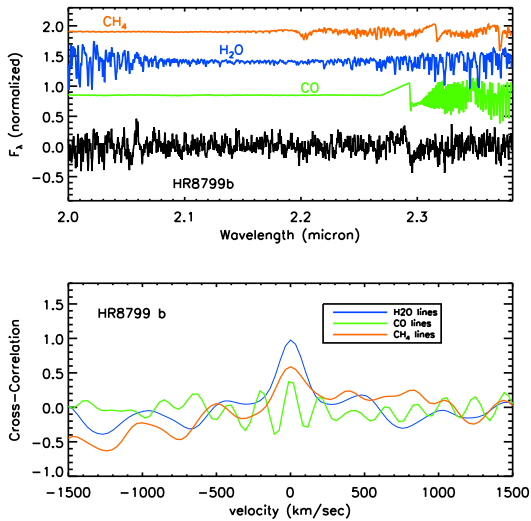}
\caption{
{\em Top}: Continuum-subtracted spectrum for \hrb.  Template spectra for
pure \methane\ (orange), \water\ (blue) and \co\ (green) are also plotted and
offset by an arbitrary amount.  {\em Bottom}: Cross-correlation functions
for \hrb\ and the template spectra plotted in the top panel.
\label{figCC}}
\end{figure}

At the full resolution and sampling provided by OSIRIS, errors in flux are
correlated across several wavelength channels and may alter the relative depth
of narrow absorption features.  Subtracting the continuum, modeled as a
smoothed version of the observed spectrum, removes many residual speckle
artifacts that vary smoothly with wavelength (K13).  Inevitably, however,
speckle artifacts that depend more strongly on wavelength (e.g., produced by
optics near the focal plane) likely remain even in a continuum-subtracted
spectrum.  Despite potential lingering artifacts, this strategy worked well in
K13 and is adopted here using the ensemble of data from 2009 through 2013,
median-combined into a single spectrum. Given the wide angular separation of
\hrb, residual speckle artifacts should be less severe than they were for \hrc.
The RMS uncertainties in the continuum-subtracted spectrum closely matches the
photon-noise limit.  The low-$R$ spectra from various observing runs show no
significant differences across this four year time frame and intrinsic
variability would likely occur over timescales of days rather than years and
impact broader wavelength ranges than the spectral lines of interest.

\section{Model Spectra}

For this study, the grid of exoplanet atmospheres models described in
B11\nocite{Barman2011a} was updated to include the methane linelist from
\cite{Yurchenko2014} supplemented with optical opacities from
\cite{Karkoschka2010}. The former update added approximately 10 billion
transitions to the overall molecular opacities and improves the accuracy of
line strengths at high temperatures --  across the $K$-band alone there are
roughly a billion transitions.  The methane abundance is low in non-equilibrium
chemistry models appropriate for \hrb\ (B11) and, consequently, there were only
minor changes to the model atmosphere structures after this opacity update.
Nevertheless, the ExoMol methane list is currently the most accurate and
complete list available and the observations analyzed here have a resolution
where accurate line data are important when studying atmospheric abundances.  

The opacity of solid and liquid particles suspended in the atmosphere (clouds)
are included using the parameterized intermediate cloud model described in
B11\nocite{Barman2011a}. Briefly, the lower boundary of the clouds are
determined by chemical equilibrium while the upper boundary is described by an
exponential decay that begins at a specified pressure. This outer pressure is a
free parameter that establishes the cloud thickness, allowing for a range of
models with cloud opacity between the high (DUSTY) and cloud-free (COND) cases
often used to bracket the importance of clouds \cite[]{Allard2001}.  The
particle size distribution is a log-normal with mean size set to 5 $\mu$m.
Cloud thickness plays an important role in determining the overall spectral
shape and is included as a free parameter in the model fits discussed below
(following B11).

Synthetic spectra were calculated with a wavelength sampling of 0.05\AA\ from
1.4 to 2.5 $\mu$m.  Each of these synthetic spectra was convolved with a
Gaussian kernel with FWHM matching the OSIRIS spectral resolution before
interpolating onto the observed wavelength grid.  These medium-resolution
spectra were continuum-subtracted following the steps in K13.

\section{Results}

\begin{figure}[t]
\plotone{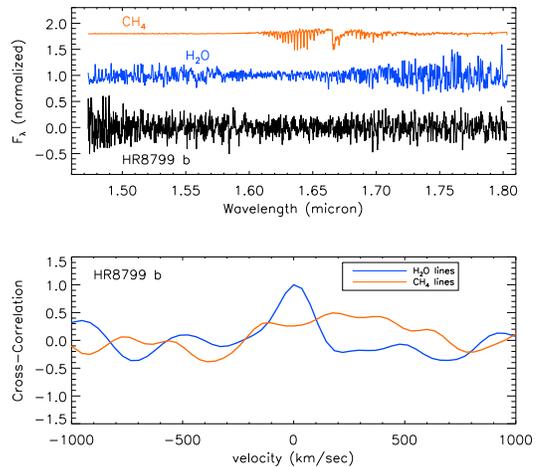}
\caption{
Same as Figure \ref{figCC}, but for $H$ band.  Only very weak \co\ lines are
present in the $H$ band and, therefore, were not searched for.  The observed
continuum-subtracted spectrum shows no correlation with the \methane\ template.
\label{figCC2}}
\end{figure}

\subsection{HR8799 b versus c}

Figure \ref{fig1} compares the $R \sim 4000$ spectrum of \hrb\ to that of
\hrc.  Both exoplanet spectra contain a similar set of absorption features,
with most of the similarities at wavelengths of prominent H$_2$O absorption;
however, most lines are deeper in the spectrum of b.  The CO (2,0) band-head at
$\sim$ 2.3 $\mu$m is detected now with medium spectral resolution but only
marginally present at low-$R$ (B11)\nocite{Barman2011a}.  This band-head
appears slightly deeper and has a shallower slope in the spectrum of b than c,
the result of stronger water absorption and additional \methane\ lines
for \hrb\ in this wavelength range.  Three regions of CH$_4$ absorption are visually
identifiable between 2.15 and 2.4 $\mu$m (lower panel of Fig. \ref{fig1}).  The
strongest methane lines are seen around 2.32 and 2.37 $\mu$m. None of these
features are present in the spectrum of c, consistent with the methane
non-detection reported by K13\nocite{Konopacky2013}.  A few additional
\methane\ lines are marginally identified around 2.25 $\mu$m, however, many of
the weaker CH$_4$ lines overlap with those of H$_2$O and CO, making
visual identification difficult.

\begin{figure}[t]
\plotone{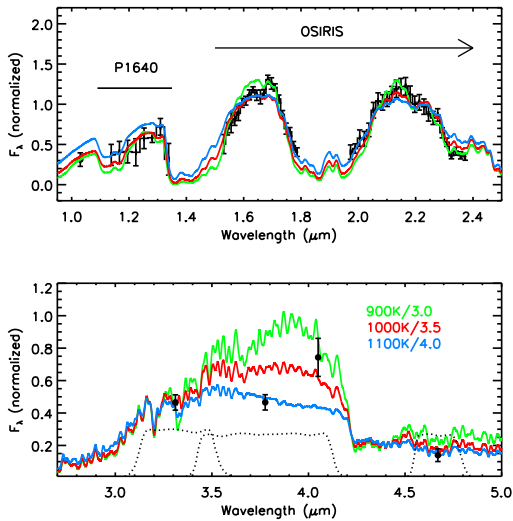}
\caption{
Example synthetic spectra (red, blue, and green) for three solar-abundance
models.  The range of model-observation agreement shown here is representative
of most models explored here.  The largest discrepancies are between 3.5 and
4.5\microns.  See text for details of the model and references for the
observations.  
\label{figmodcomp1}}
\end{figure}
\subsection{Cross-Correlation}

As an additional confirmation that spectral features from all three molecules
are present in the OSIRIS spectrum of \hrb, a cross-correlation analysis was
used following K13.  In this case, a new \methane\ template was calculated
using roughly a billion of the strongest transitions across the $K$ band
\cite[]{Yurchenko2014}.  Water and CO templates were calculated using lists
from \cite{Barber2006} and \cite{Goorvitch1994}.  Figure \ref{figCC} compares
the cross-correlation functions (CCFs) for each template.  Peaks in the CCFs
are found for all three molecules at identical velocities, centered on zero
km/s.  As was the case for \hrc, a peak with maximum near 1 is
found for H$_2$O, indicating that this molecule contributes most of the
spectral lines.  CO also shows the characteristic ringing pattern, produced by
near-repeating patterns of \co\ lines, many separated by roughly 200 km/s in
velocity. This same CCF pattern was found for CO in \hrc\ by K13, with peak
of 0.6 compared to 0.4 here.  A peak in the CCF for \methane\ is present, but
not as prominent as found for \water, mainly because much of the weak-line
information is blended with other stronger lines at this resolution.
 
$H$-band observations were previously obtained using the same telescope and
instrument combination as used for $K$-band and were originally analyzed at low
resolution in B11\nocite{Barman2011a}.  These data have been reanalyzed and
continuum-filtered at full resolution (also $R \sim 4000$) following the same
steps as $K$ band.  The average SNR of these data is about 3 times lower than
the $K$ band data and show a pattern of lines that are visually difficult to
identify.  B11 concluded that water is the dominant molecular opacity source in
this wavelength range and this is confirmed here by a cross-correlation analysis.
The CCFs for \water\ and \methane\ templates across the $H$ band are plotted in
Figure \ref{figCC2}.  Correlating the data with an $H$-band \water\ template
yields a strong peak in the CCF (Fig. \ref{figCC2}).  No peak was detected for
an $H$-band \methane\ template (using the ExoMol list).

The \hrb\ $K$-band spectrum also correlates well with the \hrc\ spectrum,
another indication that both planet spectra share similar spectral patterns of
\water\ and \co\ lines.  The CCF was also recalculated for \hrc\ using the new
\methane\ template and no peak was found, confirming the non-detection of
\methane\ reported by K13.

\subsection{Model Comparisons}

For the atmospheric abundance study presented below, it is important that the
underlying thermal structure of the models, especially across most of the
photosphere, is reasonably correct.  Atmospheres of giant planets are generally
close to local thermodynamic equilibrium and, thus, a model matching the
planet's spectral energy distribution over a wide wavelength range should have
a thermal profile that reasonably approximates the planet's average thermal
structure.  A similar model comparison as performed by B11\nocite{Barman2011a}
was repeated, but comparing updated models with the new \methane\ list to the
new medium-resolution unfiltered spectrum discussed above and additional flux
calibrated data spanning 1 to 5 \microns.  For wavelengths less than 1.8 $\mu$m
the z/Y-band flux of \cite{Currie2011}, the low-$R$ P1640 $J$-band spectrum
\cite[]{Oppenheimer2013} and the $H$-band spectrum from B11\nocite{Barman2011a}
were used.  Photometric data was used at wavelengths longer than 3 \microns\
\cite[]{Galicher2011,Skemer2012,Currie2014}.  The data used for the model fits
are plotted in Figure \ref{figmodcomp1}. 

The relative flux calibration of the near-IR spectra from OSIRIS and P1640
could impact the model comparisons, as these spectra are calibrated using $H$
and $K_s$ photometry that have a range of reported values and uncertainties
\cite[]{Marois2008, Metchev2009, Esposito2013}.  It is also possible that \hrb\
is variable as indicated by studies of brown dwarfs \cite[]{Metchev2015}.  Such
potential issues are not accounted for here and the $H$ and $K$ flux
calibration described in B11\nocite{Barman2011a} is used.  A small scaling was
applied so that the $H$-band portion of the P1640 spectrum matches the $H$-band
OSIRIS spectrum, resulting in a slight change in the absolute fluxes of the
P1640 $J$-band spectrum plotted in Figure \ref{figmodcomp1}.  

Example spectra of the best matching solar abundance models are plotted in
Figure \ref{figmodcomp1}, for a range of \teff\ between 900 and 1100K and
\logg\ between 3.0 and 4.0.  The best matching model has the
same gravity as found previously by B11 (\logg\ = 3.5) but cooler (\teff\ =1000K).
Across the near-IR, the model spectra do a reasonable job
reproducing the observations. When fitting the observations (photometric and
spectroscopic), all data are weighted equally.  Between 3 and 5\microns,
however, the model comparisons to available photometric data (from a variety of
telescopes, epochs and image processing methods) show disagreements at
$L^\prime$ and $\sim 4$ \microns.  Similar levels of disagreement at this
wavelength are seen in other model comparisons \cite[]{Marley2012, Skemer2012,
Currie2014}. These photometric observations probe similar atmospheric depths
and, therefore, pull model fits in opposite directions in both \teff\ and
\logg. Further observations across this spectral region is probably warranted.
Previous studies of \hrb\ result in a range of \teff\ between 750 and 1200K and
\logg\ between 3 and 4.5 (see Marley et al.  2012\nocite{Marley2012}, their
Table 1), encompassing the range explored here.   

Model atmosphere parameters are sometimes intentionally biased to match
predictions of cooling tracks, providing guaranteed consistency between \teff,
\logg, (and \lbol) and mass/radius expectations.  Such forced agreement can be
useful, but is not done here.  More important is finding a range of atmospheric
temperature-pressure profiles that yield pseudo-continua that are as consistent
as possible with the observed spectral energy distribution.  Good agreement
between bulk parameters found by atmosphere-only fitting and those inferred
from cooling tracks, age and luminosity can be challenging for low-temperature
planet-mass objects, likely a result of complexities of their atmospheres not
reproducible by time-independent one-dimensional models with simplistic cloud
prescriptions.  Consequently, a conservative ranges of \logg\ ($\pm$ 0.5) and
\teff\ ($\pm$ 100K) around the best matching values are used for the abundance
analysis discussed below, despite potential inconsistencies with cooling track
predictions at the high and low ends of these ranges.

\subsection{Mole Fractions of CO, CH$_4$ and H$_2$O}
\label{molefractions}

At low spectral resolution, the slopes on either side of the $K$-band peak are
very similar to those seen in substellar atmospheres where water is the
dominant opacity source.  B11\nocite{Barman2011a} showed (their Fig. 14) that
\co\ and \methane\ likely contribute to the $K$-band opacity but, at low
resolution, the \co\ band head was not confidently detected nor any significant
hints of \methane, making it difficult to assess their relative importance.
\cite{Bowler2010} also found no evidence of methane absorption in their
narrow-band spectrum covering $\sim 2.1$ to 2.2 $\mu$m. With resolved
absorption features from all three molecules identified, their individual
abundances can now be determined.

Evidence for quenching of \co\ and \methane\ in the atmospheres of the HR8799
planets has been established elsewhere \cite[]{Hinz2010, Barman2011a,
Skemer2012, Marley2012}.  An important consequence of this mixing-induced
non-equilibrium chemistry is that the mole fractions should be nearly
independent of height across the photosphere.  The absence of height-dependence
greatly simplifies the parameterization of the mole fractions, allowing
straightforward model fitting using standard \chisq\ minimization.

The mole fractions are determined sequentially by computing grids of synthetic
spectra (continuum filtered) with scaled molecular abundances for 9 different
temperature-pressure (T-P) profiles corresponding to \teff\ = 900, 1000, and
1100K and \logg\ = 3.0, 3.5, and 4.0. These temperatures and gravities bracket 
those found when comparing models to the broad wavelength data shown in Figure
\ref{figmodcomp1}.  A spectrum is calculated for each mole fraction value and
model temperature-pressure profile.  The mole fractions of CO, CH$_4$ and
H$_2$O are each scaled from 0 to 1000 relative to their non-equilibrium solar
abundance values using a uniform logarithmic sampling (resulting in 100
synthetic spectra per T-P profile).

\begin{figure}[t]
\plotone{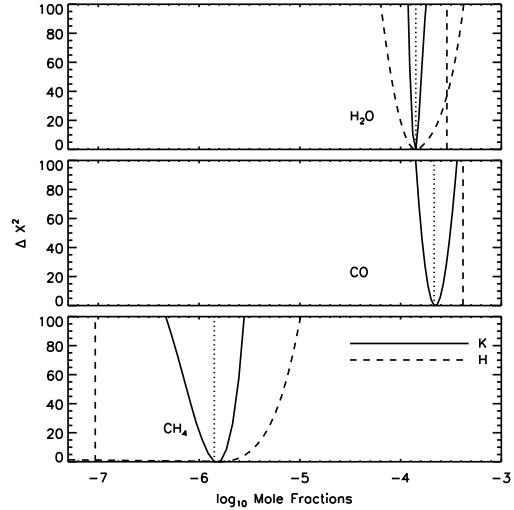}
\caption{
$\chi^2$ distributions for H$_2$O, CO and CH$_4$ mole fractions when fitting
model spectra with \teff = 1000K and gravity equal 10$^{3}$ cm sec$^{-2}$ to
the $K$ (solid) and $H$ (dashed) bands.  Vertical dotted lines indicate the
best-fit mole fractions and vertical dashed lines indicate solar
(non-equilibrium) mole fractions.  CH$_4$ was not detected in the $H$-band data,
with a 3$-\sigma$ upper limit of 10$^{-5.6}$.
\label{figMFs}}
\end{figure}

\begin{figure*}[t]
\plotone{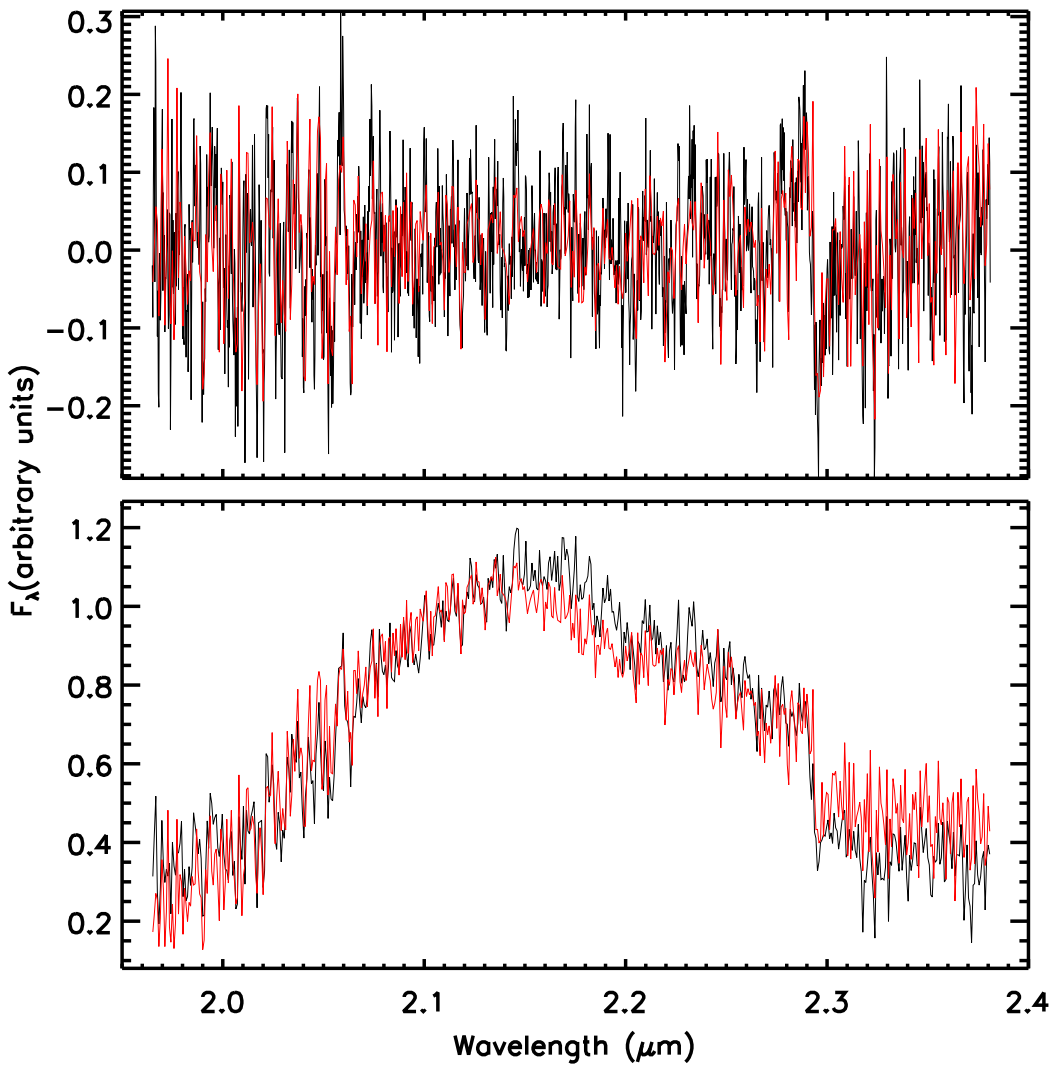}
\caption{
{\em Top}: Continuum-subtracted spectrum for \hrb\ (black) compared to 
\teff = 1000K and \logg = 3.0 model with abundances equal to those
found in Figure \ref{figMFs}.  {\em Bottom}: Spectrum of \hrb\ (black)
compared to same model plotted above but with continuum intact.
\label{figmodcomp2}}
\end{figure*}

\begin{figure}[t]
\plotone{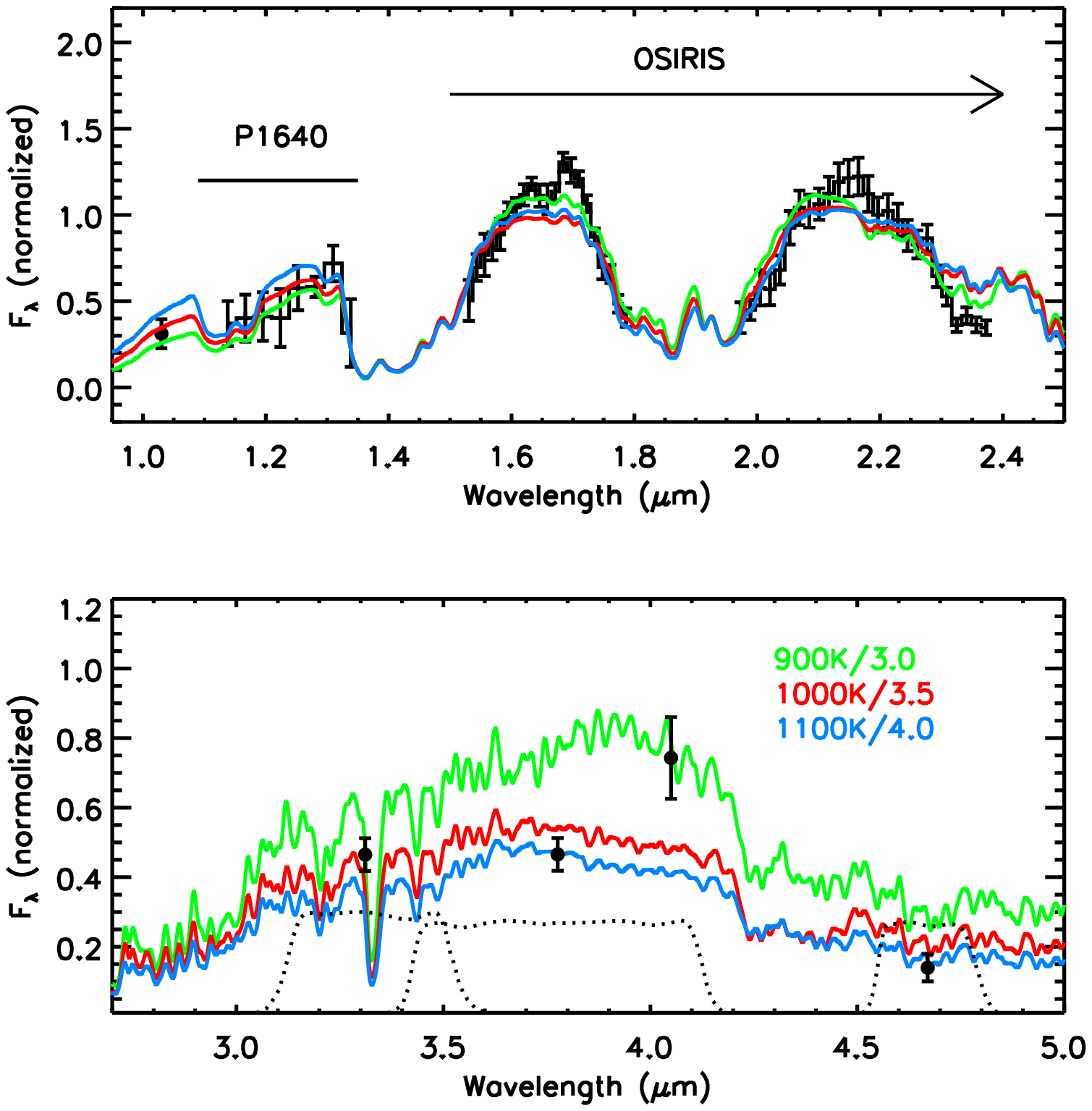}
\caption{
Same observations as plotted in Figure \ref{figmodcomp1} but compared to model
spectra with non-solar \water, \co\ and \methane\ mole fractions (see Table
\ref{tab1}).
\label{figmodcomp3}}
\end{figure}

The mole fraction of H$_2$O was determined first, keeping CO and CH$_4$ at
their solar values.  To avoid biasing in the H$_2$O mole fractions caused by
overlapping \co\ or \methane\ lines, the fit was restricted to wavelengths less
than the \co\ band head, while masking the strongest CH$_4$ line region (near
2.2 $\mu$m).  Following this step, a second grid of synthetic spectra was
calculated with the H$_2$O mole fraction equal to the best-fit value from the
previous step, CH$_4$ at its solar value, and now mole fractions of CO scaled
as previously done for \water.  In this case, only wavelengths greater than
(and including) the CO band head were fit, while masking the strong \methane\
line near 2.32 and 2.37\microns.  In the final step, using the best-fitting
H$_2$O and CO mole fractions, a third grid of synthetic spectra was made with
scaled CH$_4$ mole fractions. Only the strongest CH$_4$ lines from 2.2 to 2.37
$\mu$m were included in this fit.  

The model with the overall lowest $\chi^2$ from this three-step fitting process
has a slightly lower gravity (\logg = 3.0) than the best matching solar
abundance model compared to the full SED.  The $\chi^2$ distributions for the
mole fractions in this model are plotted in Figure \ref{figMFs}, with $\log$
mole fractions of H$_2$O, CO and CH$_4$ are $-3.85\pm 0.01$, $-3.67\pm 0.02$,
and $-5.85\pm 0.04$, respectively, with the model spectrum compared to the
$K$-band observations in Figure \ref{figmodcomp2}.  Both \water\ and \co\ are
about half the non-equilibrium solar value while \methane\ is about 15 times
lower than the non-equilibrium solar value (see vertical dashed lines in Fig.
\ref{figMFs}).  The best matching mole fractions for each \teff\ and \logg\ are
listed in Table 1, with formal 1-$\sigma$ errors determined from the $\chi^2$
distributions.  These values can be compared to the range of solar
(non-equilibrium) mole fractions by using Figure \ref{figOberg2} (see vertical
dashed line), discussed below.

The filtered $K$-band spectrum is used to determine the molecular abundances
primarily because it is a single data set from a single instrument with
uniformly characterized uncertainties.  It is important, however, to verify
that the resulting mole fractions still yield spectra that are in reasonable
agreement with the full SED.  Figure \ref{figmodcomp3} compares model spectra
to the full SED for three different \teff\ and \logg\ values and corresponding
mole fractions listed in Table 1, and the comparisons remain good across the
near-IR.  The thermal IR includes absorption bands from \methane, CO and
CO$_2$, providing a potential secondary test of the abundances.  In all cases, the
model flux at 3.3\microns, which probes a methane fundamental absorption band,
is close to the observed value. The fluxes across the CO and CO$_2$ absorption
bands between 4 and 5\microns\ changes considerably for the range of values
found here, however, the observed ground-based photometry, with measurements on
either side of the absorption bands, do not probe CO or CO$_2$ absorption well.  

As discussed above, \water\ is detected in the med-resolution $H$-band spectrum
by cross-correlation (Fig. \ref{figCC2}).  Mole fractions for \water\ were obtain by
fitting the $H$-band data in a similar manner as for $K$ band, using synthetic
spectra with scaled \water\ and \methane\ mole fractions.  A clear
\chisq-minimum was found for water for the same mole fraction as found by
fitting the $K$-band spectrum, but with much larger uncertainties.  CH$_4$ was
not detected (consistent with the cross-correlation test) and only an upper
limit was found (see dashed lines in Fig.  \ref{figMFs}).

The mole fractions determined above are sensitive to the thermal profile,
surface gravity, individual line broadening parameters, and potential residual
artifacts in the data.  Uncertainties associated with these properties are
difficult to quantify without a more complex statistical analysis of the model
comparisons and is beyond the intended scope of this paper.  For example, the
mole fractions depend on the model atmosphere surface gravity, but it is
unlikely the planet's gravity falls outside the range explored here, given the
limits placed on gravity by other data. The details of the cloud properties
(coverage and thickness) are partially mitigated by the continuum subtraction
process but may still impact the inferred mole fractions indirectly though
degeneracies in the surface gravity and effective temperature estimation.
Uncertainties in \teff\ translate into a small ($\sim 0.1$ dec) changes in the
mole fractions.  Future refinement of the mole fractions will benefit most by
improved determinations of \logg.  Surface gravity broadens spectral lines and,
in principle, the width of auto-correlation functions for models of various
surface gravities could be used to determine gravity's contribution to the
total line broadening.  Unfortunately, even a resolution of 4000 ($\sim 75$ km
s$^{-1}$) is too low for such an exercise to yield better gravity estimates
than those based on model fits to lower resolution data (e.g., fitting shapes
of $H$ and $K$ band spectra).

\begin{deluxetable}{llccc}
\tablecolumns{5}
\tablecaption{$\log$ Mole Fractions}
\tablehead{&  & &\colhead{\logg} &    \\
\colhead{molecule}& \colhead{\teff} & \colhead{3.0}&\colhead{3.5} & \colhead{4.0}}
\startdata
                 & 900        & -3.82$\pm 0.01$ & -3.39$\pm 0.01$ & -3.09$\pm 0.01$\\
\water           & 1000       & -3.85$\pm 0.01$ & -3.55$\pm 0.01$ & -3.12$\pm 0.01$\\
                 & 1100       & -3.91$\pm 0.01$ & -3.58$\pm 0.01$ & -3.24$\pm 0.01$\\
                 &            &               &               &              \\
                 & 900        & -3.61$\pm 0.02$ & -3.48$\pm 0.02$ & -3.30$\pm 0.02$\\
\co              & 1000       & -3.67$\pm 0.02$ & -3.61$\pm 0.02$ & -3.33$\pm 0.02$\\
                 & 1100       & -3.72$\pm 0.02$ & -3.61$\pm 0.02$ & -3.30$\pm 0.02$\\
                 &            &               &               &              \\
                 & 900        & -5.85$\pm 0.04$ & -5.55$\pm 0.04$ & -5.18$\pm 0.04$\\
\methane         & 1000       & -5.85$\pm 0.04$ & -5.48$\pm 0.04$ & -5.12$\pm 0.04$\\
                 & 1100       & -5.79$\pm 0.04$ & -5.42$\pm 0.04$ & -5.06$\pm 0.04$\\
\enddata
\label{tab1}
\end{deluxetable}

\subsection{Estimating $K_{zz}$}

Vertical mixing is often characterized by the coefficient for eddy diffusion
(\kzz) where the vertical mixing timescale above the convection zone is $L_{\rm
eff}$/\kzz\ and $L_{\rm eff}$ is an effective length scale usually a few tenths
the pressure scale height\cite[]{Smith1998}.   Below the point in the atmosphere where the mixing
timescales are shorter than chemical reaction timescales, the atmosphere will be in
chemical equilibrium.  Most chemical reactions are fast, however, the reactions
governing \co\ and \methane\ have one or more rate-limiting steps that result
in long chemical timescales \cite[]{Visscher2010, Zahnle2014} that rapidly
increase with decreasing density, quickly exceeding the age of the planet by
many orders of magnitude.  The ultimate consequence is that, for plausible
values of \kzz, the photospheric mole fractions of \co\ and \methane\ (as well
as others, e.g., N$_2$, NH$_3$, and CO$_2$) may no longer depend on the
photosphere conditions but instead on the conditions deeper in the atmosphere
where the chemical and mixing timescales become comparable.  

\begin{figure}[t]
\plotone{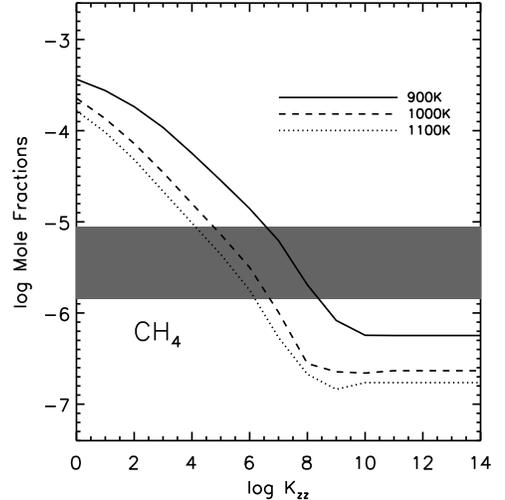}
\caption{{\em Top}:
Photospheric mole fractions for methane as functions of the coefficient for
eddy diffusion in models with \logg\ = 3.5 and \teff\ = 900, 1000, and 1100K
(solid, dashed and dotted lines, respectively).  The horizontal bar indicates
the range of methane mixing ratios inferred from the $K$-band spectrum (Table
1).  \label{figKzz}}
\end{figure}

The detection of \co\ in Jupiter's atmosphere is an excellent example of
vertical mixing in action.  For the very low temperatures in Jupiter's
atmosphere ($\sim 100$K), the majority of carbon is in \methane\ at an
abundance that is essentially constant with height. Therefore, if quenching
occurs in the atmosphere, the photospheric \methane\ will be unchanged.  The
equilibrium photospheric mole fractions of CO, on the other hand, are small and
rapidly decreasing with height. If pure chemical equilibrium persisted
throughout, \co\ would be nearly impossible to observe.   Yet, photospheric
\co\ absorption has been measured, evidence that non-equilibrium chemistry is at work, and
the \co\ abundance used to estimate the value of \kzz\
\cite[]{FegleyLodders1994}.   

\begin{figure*}[t]
\plotone{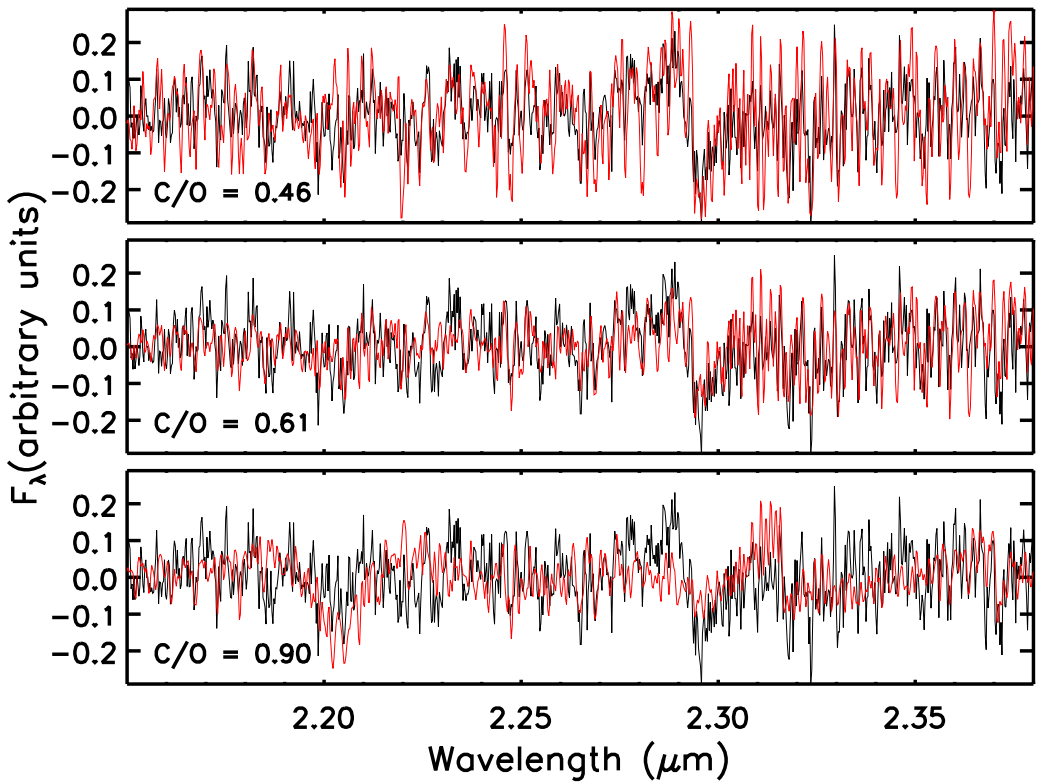}
\caption{
Continuum-subtracted spectrum for \hrb\ (black) compared to models (red) with low
(top), high (bottom) and best-matching (middle) carbon-to-oxygen (C/O) ratios
found in \S \ref{sec4.6}.  At low C/O, \water\ and \co\ absorption lines are
stronger than observed, while \methane\ absorption lines are weaker.  At high
C/O, \methane\ absorption is stronger than observed while \water\ and \co\ are
too weak.  The short-wavelength portion of $K$-band is not shown for clarity,
though the general level of agreement is comparable to what
is plotted.
\label{figmodcomp4}}
\end{figure*}

In the atmospheres of young giant exoplanets as hot as \hrb\ ($\sim 1000K$)
the situation for \co\ and \methane\ is reversed \cite[]{Barman2011b,
Zahnle2014}.  At pressures (between 1 and 10 bar,
the temperatures are hot enough for \co\ to be thermochemically favored
over \methane.   Also, the quenching can occur at or near depths where
equilibrium mole fractions are height-{\em dependent} for \methane\ and
nearly height-{\em independent} for \co.  An important outcome is that, much like \co\ in
Jupiter, measurements of \methane\ can be used estimate \kzz\ in the
atmospheres of planets like \hrb.

The photospheric \methane\ mole fractions versus \kzz\ are plotted in Figure
\ref{figKzz} for solar composition and a range of \teff\ appropriate for \hrb.
Only for $\log(K_{zz}) > 10$ does \kzz\ become independent of \methane.  The
range of inferred mole fractions for methane (horizontal shaded region in Fig.
\ref{figKzz}) indicate $\log(K_{zz})$ around 7 for \teff$ = 1000K$; however,
the mole fractions at the quenching depth are temperature and gravity dependent
leading to an uncertainty in \kzz\ of about 100 cm$^2$ s$^{-1}$.  Large
departures from solar C and O abundances could change the inferred \kzz,
however, as argued below, it is unlikely that C/O is very far from solar.

\subsection{C/O Ratio}
\label{sec4.6}

The C/O ratio for \hrc\ was found to be slightly above that of the host star,
tentatively favoring the core-accretion formation scenario over gravitational
instability (K13).  Determining the C/O for \hrb\ is an important next step in
understanding the formation history of this planetary system.

The photosphere of \hrb\ is not in chemical equilibrium, as discussed above,
with both \co\ and \water\ quenched deep in the atmospheres.  Deep quenching of
carbon and oxygen-bearing molecules has important implications for inferring
the relative C and O element abundances from the molecular mole fractions.
Even young giant planets, still hot from recent formation, have atmospheres
cool enough to allow solids and liquids to form, including silicate grains
(e.g., MgSiO$_3$, Mg$_2$SiO$_4$). The abundance of these grains is limited by
the overall abundance of Si, and will sequester a non-negligible fraction of
the Oxygen atoms.  The photosphere of \hrb\ is well below the temperatures
needed for condensation to occur and, as a result, the inventory of Oxygen
should account for both the mole fractions of silicate grains and
Oxygen-bearing molecules. However, for the atmosphere of \hrb, models indicate
that the temperature at the quenching depths is above 2000K and, thus, above
the condensation temperature for major oxygen-depleting grains.  In the case of
the HR8799 planets, non-equilibrium chemistry simplifies the C/O ratio
dependence on atmospheric mole fractions (N) to
\begin{equation}
\frac{C}{O} = \frac{N(CH_4)+N(CO)}{N(H_2O)+N(CO)},
\end{equation}
\noindent and for small amounts of \methane, the C/O ratio is determined by \water\ and
\co\ alone.

Equation 1 and the mole fractions determined above (Table 1) for each model
atmosphere \teff\ and \logg, results in C/O values between 0.4 and 0.7.  This
large range in C/O is mostly driven by the comparably large range of acceptable
values for \logg, with C/O decreasing as \logg\ increases. The mole fractions
for the best matching model yield C/O = 0.61 $\pm 0.05$ (with 1-$\sigma$ error
determined from the formal errors in Table 1).  Figure \ref{figmodcomp4}
compares the continuum-filtered observations to the best overall model from
\S \ref{sec4.6} (C/O = 0.61) and models with high and low C/O, to
illustrate the changes in the spectrum as the relative molecular abundances
change.

\begin{figure}[t]
\plotone{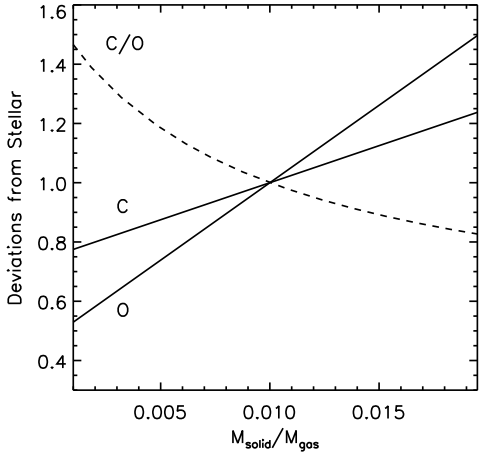}
\caption{
Carbon and Oxygen abundances (solid lines) relative to stellar values, along
with the corresponding C/O ratio (dashed line), using the \cite{Oberg2011}
model for planets forming between the \water\ and CO$_2$ frostlines.  The disk
is assumed to have the same grain/gas fraction as the interstellar medium
(0.01). See \cite{Oberg2011} for the specific details of their model.  
\label{figObergCO}}
\end{figure}

\section{Atmospheric Composition and Formation}
\label{sec5}
If giant planets form primarily by a quick one-step process via gravitational
instabilities (GI), their atmospheres should have element abundances equal to
the host star \cite[]{Helled2009}. On the other hand, if giant planets form
primarily by the multi-step core-accretion (CA) process, a range of element
abundances are possible \cite[]{Oberg2011}.   The abundances of a gas giant's
atmosphere formed via CA primarily depend on the location of formation relative
to the frostlines for major carbon and oxygen bearing molecules in the disk
(namely \water, CO$_2$ and \co) and the amount of solids acquired by the planet
during the runaway accretion phase.  The four planets orbiting HR8799 offer an
excellent opportunity to test this idea.  Each planet currently orbits between
the \water\ and CO$_2$ frostlines and potentially built up atmospheres from gas
with similar amounts of solids.

In K13, the observed continuum-filtered spectrum of \hrc\ was compared to
atmosphere models restricted to a sequence of C and O element abundances
derived from the \cite{Oberg2011} chemical model.  The {\"O}berg et al.  model
provides values for the C and O abundances, relative to the stellar values, for
different amounts of solid accretion during the buildup of the planetary
envelope.  These abundances are plotted in Figure \ref{figObergCO} for planets
forming between the \water\ and CO$_2$ frostlines.  In this model, both C and O
abundances are linear functions of solid accretion, M$_{\rm solid}$/M$_{\rm
gas}$, with slope and intercept set by the fraction of C (or O) sequestered by
condensate formation and the overall grain/gas fraction in the disk.  These
assumptions are based on observations of protoplanetary disks and the
interstellar medium (see Table 1 of {\"O}berg et al. 2011).  Solar C and O
abundances have been suggested for HR8799 \cite[]{Sadakane2006} and are adopted as
the baseline here \cite[]{Asplund2009}.  It should be noted, however, that this star is a
$\lambda$-boo type star with solar C, N and O abundances but sub-solar Fe-peak
elements. 

\begin{figure}[t]
\plotone{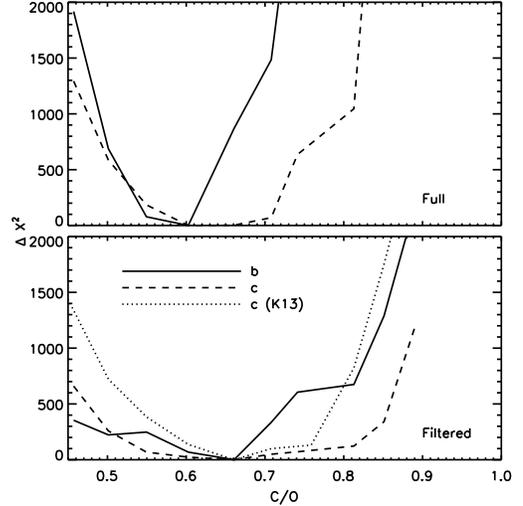}
\caption{
{\em Top}: Distribution of $\Delta \chi^2$ for models with C and O abundances
following \cite{Oberg2011} compared to the full $K$-band spectrum with flux
calibrated continuum for \hrb\ (solid line) and \hrc\ (dashed line).
{\em Bottom}: Same as top panel but for comparisons to the filtered
(continuum-subtracted) $K$-band spectrum.  The dotted line is the $\Delta
\chi^2$ distribution for \hrc\ from K13 and the dashed line is the distribution
for a reanalysis discussed in the text.
\label{figOberg1}}
\end{figure}

A high level of solid accretion during planet formation raises both C and O abundances, with O
abundances increasing more rapidly than C for the simple reason that \water-ice
is the most abundant solid between the \water\ and CO$_2$ frostlines. The
combined effect is C/O decreasing as M$_{\rm solid}$/M$_{\rm gas}$ increases.  The model
proposed by {\"O}berg et al. is a simple prescription for a complex process
and, consequently, deviations from this model are to be expected.  Despite its
simplicity, the predicted C and O abundances provide an ideal baseline for
testing potential outcomes of CA formation specific to the HR8799 system.

In order to make a direct comparison to the K13 results, the observed spectrum
of \hrb\ was analyzed in a similar manner as \hrc.  Given the \methane\ update
made to the model atmospheres, the fit was repeated for \hrc. Only the C/O
values for atmosphere accretion occurring between the \water\ and CO$_2$
frostlines (the current locations of all four HR8799 planets) were used (see
Fig. \ref{figObergCO}).  A $\chi^2$ was calculated for each
continuum-subtracted synthetic spectrum in the \teff\ and \logg\ range
described above. The resulting $\chi^2$ distributions are plotted in Figure
\ref{figOberg1}.  The best-matching C/O for \hrb\ is 0.66$^{+0.04}_{-0.08}$ and
the revised best-matching C/O for \hrc\ is 0.64$^{+0.14}_{-0.11}$, closely
matching K13, but with a broader \chisq\ distribution.  The new ExoMol methane
absorption strengths across the $K$-band are lower than those from the
\cite{Warmbier2009} linelist used in the K13 analysis resulting in smaller
\chisq\ values for larger C/O and, hence, a broader distribution of \chisq. 
 
The $K$-band continuum shape can also be used to estimate the relative
abundance of \water\ and \co.  Given the wider angular separation of b than c
(1.7 versus 1\arcsec) from the star, the continuum of b is less affected by
residual speckles.  The observed spectrum, with continuum intact, was compared
to the same set of atmosphere models and C and O abundances.  The \chisq\
distribution from this comparison is plotted in the top panel of Fig.
\ref{figOberg1} and shows both planets having C/O between $\sim 0.55$ and 0.7,
for similar $\Delta \chi^2$ as in the continuum filtered comparison.

\begin{figure}[t]
\plotone{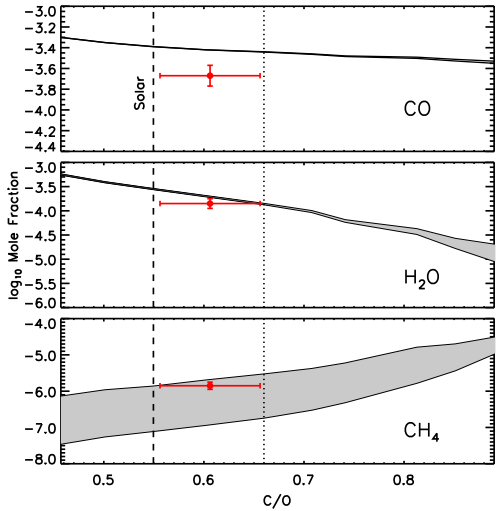}
\caption{
Photospheric mole fractions of \co, \water, and \methane\ as functions of C/O
for atmosphere models with C and O abundances as plotted in Fig.
\ref{figObergCO}.   For each molecule and C/O, a range of model values is
possible given the range of effective temperature and gravity explored here
(900 -- 1100K, 3.0 -- 4.0).  The ranges of mole fractions are plotted as shaded
regions for \water\ and \methane.  The range of CO mole fractions is narrow and
plotted as a black line with varying thickness.  These regions correspond to
the photospheric molecular abundances specifically for models of \hrb, and the
assumptions made here about vertical mixing of CO and \methane.  The
best-matching model with {\"O}berg et al. based abundances has a C/O $\sim$
0.66 (see \S  \ref{sec5}) and indicated by the vertical dotted line. The full
range of allowed C/O from \S \ref{sec5} is not shown to avoid cluttering the
Figure, but extends from 0.54 to 0.7. The intersections of this dotted line and
the shaded regions indicate the molecular mole fractions at the photosphere of
the best-fitting model.  For comparison, solar abundances and C/O are indicated
by the vertical dashed line.  The mole fractions and best matching range of C/O
determined in \S \ref{sec4.6} are plotted as red symbols with horizontal and
vertical error-bars.  The different model fitting procedures result in
consistent mole fractions for \water\ and \methane, while the scaling procedure
results in \co\ lower than in all of the model atmospheres with {\"O}berg et
al.  based C and O abundances. 
\label{figOberg2}}
\end{figure}

The mole fractions of \water, \co\ and \methane\ in the photosphere of the
best-matching CA-specific atmosphere model can be compared to those found in
\S \ref{molefractions}, where the molecular abundances were determined
independent of any formation model.  The mole fractions for the CA-specific
atmosphere models are plotted in Figure \ref{figOberg2} as functions of C/O,
for the range of \teff\ and \logg\ explored here.  Solar (non-equilibrium) mole
fractions and solar C/O are indicated in this Figure by the vertical dashed line.  The
CA-specific atmosphere model with C/O $\sim 0.66$ has \water\ and \methane\
mole fractions that closely match those found in \S \ref{molefractions}. These
values are plotted as red symbols with error bars estimated from the formal
uncertainties in Table 1 and the CO abundances corresponding to the 1$-\sigma$
range in C/O.  CO, however, is higher than found in \S \ref{molefractions} for
all C/O explored in the CA-specific atmospheres.  Even if \logg\ and \teff\ are
varied (see Table 1), the CO found by freely scaling the abundances remains
below what is predicted in the CA-specific atmospheres.  Lower \kzz\ would
lower \co, but doing so would quickly increase \methane\ to values well above
what is observed.  A lower \co\ could indicate additional chemical processes
not accounted for in the atmosphere or formation models.  The significance of
this CO difference is hard to assess given the various source of uncertainty in
the models not accounted for in the formal mole fraction error-bars.  It is
possible, for example, that the CO line broadening in the atmosphere models is
under (or over) estimated, resulting in systematic errors in the inferred
abundances (this will be investigated in a future paper).

Compared to \water\ and \methane, \co\ has the weakest dependence on C/O.
\methane\ and \water, however, are fairly sensitive to C/O, but their mole
fractions can be small ($< 10^{-4}$) for super-solar C/O.  Measuring C/O from
CO alone would require a CO abundance more precisely derived than most data
presently allow.  Perhaps of greater interest, however, are the ratios of the
various molecules as C/O changes.  For example, for all but the smallest C/O,
\co/\water\ should be greater than 1.  Only for low surface gravity ($< 3.5$)
do the model fits from \S \ref{molefractions} result in N(\co) $>$ N(\water)
and, as gravity increase, the discrepancy between the predicted and observed
\co\ and \water\ abundances grows. This may provide indirect evidence
supporting low surface gravity for \hrb.  
 
\section{Summary and Conclusions}

A new $K$-band spectrum of \hrb, with spectral resolution of $\sim 4000$, was
measured and simultaneous detections of water, carbon monoxide and methane
absorption lines were made.  Identification of lines from each molecule is
possible by eye and confirmed by cross-correlating the observed spectrum with
absorption templates for each molecule.  The abundance of each molecule was
determined by a fitting procedure where the mole fractions are treated as free
parameters in sequences of synthetic spectra.  The ensemble of near-IR to
thermal IR observations for \hrb\ are best reproduced by an atmosphere with
\teff = 1000, \logg\ = 3.5.  The $\log$ mole fractions are found to be between
-3.09 and -3.91 for \water, between -3.30 and -3.72 for \co, and between -5.06
and -5.85 for \methane. The best matching models have C/O between 0.55 and 0.7
depending heavily on \logg, as also found in K13, with C/O decreasing with
increasing \logg.  If the surface gravity is closer to \logg\ = 4, as predicted
by hot-start cooling tracks \cite[]{Baraffe2003}, then \co\ would be less
abundant than \water\ (see Table 1) with a C/O $\sim 0.4$.  Such a situation
would require C and O abundances that deviate from those predicted by the
\cite{Oberg2011} model, perhaps resulting from planetesimal accretion or core
dredging.  

A cross-correlation analysis of $H$-band data from B11, taken at a similar
spectral resolution as the $K$-band data, revealed no methane signature.  A
detection of \methane\ absorption lines in the $K$-band but not in $H$ may seem
unexpected when compared to late-type brown dwarfs that often show more
prominent \methane\ absorption in $H$ than $K$.   As shown here, the methane
abundance in \hrb\ is many orders of magnitude below that found in the
atmospheres of field brown dwarfs, resulting in weaker methane absorption at
all wavelengths.  For photospheric temperatures and pressures appropriate for
\hrb, the average methane opacity (in units of cm$^2$/molecule) across $K$ is
about 10 times stronger than the average opacity across $H$, while the average
water opacity is about 10 times stronger than methane in $H$ and 10 times
weaker than methane\ in $K$.  Consequently, water opacity likely overwhelms
that of methane across the $H$ band. The characteristics of these opacity
sources, combined with lower SNR in the $H$-band data, all contribute to the
non-detection of methane in the $H$-band spectrum. 
 
Low methane abundance in \hrb\ is consistent with previous studies of
non-equilibrium chemistry in hot, low gravity atmospheres.  The mixing ratio of
\methane\ in young giant exoplanets is, potentially, a useful probe of the
vertical mixing.  Using the inferred \methane\ abundance, the coefficient of
eddy diffusion was found to be greater than 10$^6$ and very near 10$^7$
cm$^2$/s for the preferred \teff\ = 1000K. These \kzz\ values are consistent with
values appropriate for Jupiter's deep troposphere \cite[]{Visscher2010}.
\cite{Zahnle2014} explore the topic of non-equilibrium \methane\ in the context
of young giant planets and their predictions are in good agreement with the
values presented here.

\hrb\ and c are the first exoplanets to have C/O ratios determined from
spectroscopic data with high SNR and high spectral resolution ($R \sim 4000$).
Young giant planets, still orbiting their star beyond the \water-frostline,
provide important opportunities to study the link between formation and
atmospheric chemistry.  The C/O ratios of both \hrb\ and c are similar and
potentially super-stellar, but stellar values are not completely excluded.  
Refinements of C/O and individual molecular abundances will require
improvements in the determination of surface gravity, independent of
evolutionary models, perhaps with near-IR observations at even higher spectral
resolution.  Observations with the {\em Hubble Space Telescope}, at wavelengths
within near-IR water absorption bands, might also help narrow the range of
allowed abundances found here (Rajan et al., in prep). But comparisons to such
data would face similar limitations as here, associate with effective
temperature and gravity.  More observations across the thermal infrared (3 to
5\microns), perhaps with future ground-based integral field spectrographs or
with the {\em James Webb Space Telescope}, could provide more complete coverage
of strong \methane, CO and CO$_2$ absorption bands. New measurements across
this wavelength range would also help improve effective temperature and gravity
estimates.  Model uncertainties need to be better understood as well, in
particular those associated with clouds and natural molecular line broadening.

There are now three young directly imaged planet-mass companions (\hrb, c and
2M1207B) that show evidence of disequilibrium chemistry in their deep
troposphere.  While three is too few to constitute a trend, the data
and models suggest that chemical quenching is common among this class of
exoplanet.  For planets as warm as those orbiting HR8799, the quenching likely
occurs below the condensation depths and, consequently, bulk C/O ratios are
inferable form their photospheric \water\ and \co\ abundances.  Given the broad
absorption features of these molecules, it may be possible to infer C/O for
massive planets from the low resolution spectra provided by GPI, P1640, and
SPHERE.

\acknowledgements
We thank the referee, Thayne Currie, for useful comments and a careful review
of this paper.  We also thank Sergei Yurchenko and the ExoMol group for
providing a copy of their methane linelist in advance of publication.  We also
thank Peter Hauschildt, Isabelle Baraffe, Gilles Chabrier and Mark Marley for
fruitful discussions during the course of this work.  The data presented herein
were obtained at the W.M. Keck Observatory, operated as a scientific
partnership among the California Institute of Technology, the University of
California and the National Aeronautics and Space Administration.  The
Observatory was made possible by the generous financial support of the W.M.
Keck Foundation. The authors wish to recognize and acknowledge the very
significant cultural role and reverence that the summit of Mauna Kea has always
had within the indigenous Hawaiian community.  We are most fortunate to have
the opportunity to conduct observations from this mountain.  Most of the
numerical work was carried out at the NASA Advanced Supercomputing facilities.
This research was support by the NSF and NASA grants to LLNL and the University
of Arizona.  This research was also support by JPL/NexSci RSA awards.  We thank
all these institutions for their support.


\begin{thebibliography}{39}
\expandafter\ifx\csname natexlab\endcsname\relax\def\natexlab#1{#1}\fi

\bibitem[{{Allard} {et~al.}(2001){Allard}, {Hauschildt}, {Alexander},
  {Tamanai}, \& {Schweitzer}}]{Allard2001}
{Allard}, F., {Hauschildt}, P.~H., {Alexander}, D.~R., {Tamanai}, A., \&
  {Schweitzer}, A. 2001, \apj, 556, 357

\bibitem[{{Asplund} {et~al.}(2009){Asplund}, {Grevesse}, {Sauval}, \&
  {Scott}}]{Asplund2009}
{Asplund}, M., {Grevesse}, N., {Sauval}, A.~J., \& {Scott}, P. 2009, \araa, 47,
  481

\bibitem[{{Baraffe} {et~al.}(2003){Baraffe}, {Chabrier}, {Barman}, {Allard}, \&
  {Hauschildt}}]{Baraffe2003}
{Baraffe}, I., {Chabrier}, G., {Barman}, T.~S., {Allard}, F., \& {Hauschildt},
  P.~H. 2003, \aap, 402, 701

\bibitem[{{Barber} {et~al.}(2006){Barber}, {Tennyson}, {Harris}, \&
  {Tolchenov}}]{Barber2006}
{Barber}, R.~J., {Tennyson}, J., {Harris}, G.~J., \& {Tolchenov}, R.~N. 2006,
  \mnras, 368, 1087

\bibitem[{{Barman} {et~al.}(2011{\natexlab{a}}){Barman}, {Macintosh},
  {Konopacky}, \& {Marois}}]{Barman2011a}
{Barman}, T.~S., {Macintosh}, B., {Konopacky}, Q.~M., \& {Marois}, C.
  2011{\natexlab{a}}, \apj, 733, 65

\bibitem[{{Barman} {et~al.}(2011{\natexlab{b}}){Barman}, {Macintosh},
  {Konopacky}, \& {Marois}}]{Barman2011b}
---. 2011{\natexlab{b}}, \apjl, 735, L39

\bibitem[{{Bowler} {et~al.}(2010){Bowler}, {Liu}, {Dupuy}, \&
  {Cushing}}]{Bowler2010}
{Bowler}, B.~P., {Liu}, M.~C., {Dupuy}, T.~J., \& {Cushing}, M.~C. 2010, \apj,
  723, 850

\bibitem[{{Currie} {et~al.}(2014){Currie}, {Burrows}, {Girard}, {Cloutier},
  {Fukagawa}, {Sorahana}, {Kuchner}, {Kenyon}, {Madhusudhan}, {Itoh},
  {Jayawardhana}, {Matsumura}, \& {Pyo}}]{Currie2014}
{Currie}, T., {Burrows}, A., {Girard}, J.~H., {Cloutier}, R., {Fukagawa}, M.,
  {Sorahana}, S., {Kuchner}, M., {Kenyon}, S.~J., {Madhusudhan}, N., {Itoh},
  Y., {Jayawardhana}, R., {Matsumura}, S., \& {Pyo}, T.-S. 2014, \apj, 795, 133

\bibitem[{{Currie} {et~al.}(2011){Currie}, {Burrows}, {Itoh}, {Matsumura},
  {Fukagawa}, {Apai}, {Madhusudhan}, {Hinz}, {Rodigas}, {Kasper}, {Pyo}, \&
  {Ogino}}]{Currie2011}
{Currie}, T., {Burrows}, A., {Itoh}, Y., {Matsumura}, S., {Fukagawa}, M.,
  {Apai}, D., {Madhusudhan}, N., {Hinz}, P.~M., {Rodigas}, T.~J., {Kasper}, M.,
  {Pyo}, T., \& {Ogino}, S. 2011, \apj, 729, 128

\bibitem[{{Esposito} {et~al.}(2013){Esposito}, {Mesa}, {Skemer}, {Arcidiacono},
  {Claudi}, {Desidera}, {Gratton}, {Mannucci}, {Marzari}, {Masciadri}, {Close},
  {Hinz}, {Kulesa}, {McCarthy}, {Males}, {Agapito}, {Argomedo}, {Boutsia},
  {Briguglio}, {Brusa}, {Busoni}, {Cresci}, {Fini}, {Fontana}, {Guerra},
  {Hill}, {Miller}, {Paris}, {Pinna}, {Puglisi}, {Quiros-Pacheco}, {Riccardi},
  {Stefanini}, {Testa}, {Xompero}, \& {Woodward}}]{Esposito2013}
{Esposito}, S., {Mesa}, D., {Skemer}, A., {Arcidiacono}, C., {Claudi}, R.~U.,
  {Desidera}, S., {Gratton}, R., {Mannucci}, F., {Marzari}, F., {Masciadri},
  E., {Close}, L., {Hinz}, P., {Kulesa}, C., {McCarthy}, D., {Males}, J.,
  {Agapito}, G., {Argomedo}, J., {Boutsia}, K., {Briguglio}, R., {Brusa}, G.,
  {Busoni}, L., {Cresci}, G., {Fini}, L., {Fontana}, A., {Guerra}, J.~C.,
  {Hill}, J.~M., {Miller}, D., {Paris}, D., {Pinna}, E., {Puglisi}, A.,
  {Quiros-Pacheco}, F., {Riccardi}, A., {Stefanini}, P., {Testa}, V.,
  {Xompero}, M., \& {Woodward}, C. 2013, \aap, 549, A52

\bibitem[{{Fabrycky} \& {Murray-Clay}(2010)}]{Fabrycky2010}
{Fabrycky}, D.~C. \& {Murray-Clay}, R.~A. 2010, \apj, 710, 1408

\bibitem[{{Fegley} \& {Lodders}(1994)}]{FegleyLodders1994}
{Fegley}, Jr., B. \& {Lodders}, K. 1994, \icarus, 110, 117

\bibitem[{{Galicher} {et~al.}(2011){Galicher}, {Marois}, {Macintosh}, {Barman},
  \& {Konopacky}}]{Galicher2011}
{Galicher}, R., {Marois}, C., {Macintosh}, B., {Barman}, T., \& {Konopacky}, Q.
  2011, \apjl, 739, L41

\bibitem[{{Goorvitch}(1994)}]{Goorvitch1994}
{Goorvitch}, D. 1994, \apjs, 95, 535

\bibitem[{{Helled} \& {Schubert}(2009)}]{Helled2009}
{Helled}, R. \& {Schubert}, G. 2009, \apj, 697, 1256

\bibitem[{{Hinz} {et~al.}(2010){Hinz}, {Rodigas}, {Kenworthy}, {Sivanandam},
  {Heinze}, {Mamajek}, \& {Meyer}}]{Hinz2010}
{Hinz}, P.~M., {Rodigas}, T.~J., {Kenworthy}, M.~A., {Sivanandam}, S.,
  {Heinze}, A.~N., {Mamajek}, E.~E., \& {Meyer}, M.~R. 2010, \apj, 716, 417

\bibitem[{{Ingraham} {et~al.}(2014){Ingraham}, {Marley}, {Saumon}, {Marois},
  {Macintosh}, {Barman}, {Bauman}, {Burrows}, {Chilcote}, {De Rosa}, {Dillon},
  {Doyon}, {Dunn}, {Erikson}, {Fitzgerald}, {Gavel}, {Goodsell}, {Graham},
  {Hartung}, {Hibon}, {Kalas}, {Konopacky}, {Larkin}, {Maire}, {Marchis},
  {McBride}, {Millar-Blanchaer}, {Morzinski}, {Norton}, {Oppenheimer},
  {Palmer}, {Patience}, {Perrin}, {Poyneer}, {Pueyo}, {Rantakyr{\"o}},
  {Sadakuni}, {Saddlemyer}, {Savransky}, {Soummer}, {Sivaramakrishnan}, {Song},
  {Thomas}, {Wallace}, {Wiktorowicz}, \& {Wolff}}]{Ingraham2014A}
{Ingraham}, P., {Marley}, M.~S., {Saumon}, D., {Marois}, C., {Macintosh}, B.,
  {Barman}, T., {Bauman}, B., {Burrows}, A., {Chilcote}, J.~K., {De Rosa},
  R.~J., {Dillon}, D., {Doyon}, R., {Dunn}, J., {Erikson}, D., {Fitzgerald},
  M.~P., {Gavel}, D., {Goodsell}, S.~J., {Graham}, J.~R., {Hartung}, M.,
  {Hibon}, P., {Kalas}, P.~G., {Konopacky}, Q., {Larkin}, J.~A., {Maire}, J.,
  {Marchis}, F., {McBride}, J., {Millar-Blanchaer}, M., {Morzinski}, K.~M.,
  {Norton}, A., {Oppenheimer}, R., {Palmer}, D.~W., {Patience}, J., {Perrin},
  M.~D., {Poyneer}, L.~A., {Pueyo}, L., {Rantakyr{\"o}}, F., {Sadakuni}, N.,
  {Saddlemyer}, L., {Savransky}, D., {Soummer}, R., {Sivaramakrishnan}, A.,
  {Song}, I., {Thomas}, S., {Wallace}, J.~K., {Wiktorowicz}, S.~J., \& {Wolff},
  S.~G. 2014, \apjl, 794, L15

\bibitem[{{Karkoschka} \& {Tomasko}(2010)}]{Karkoschka2010}
{Karkoschka}, E. \& {Tomasko}, M.~G. 2010, \icarus, 205, 674

\bibitem[{{Konopacky} {et~al.}(2013){Konopacky}, {Barman}, {Macintosh}, \&
  {Marois}}]{Konopacky2013}
{Konopacky}, Q.~M., {Barman}, T.~S., {Macintosh}, B.~A., \& {Marois}, C. 2013,
  Science, 339, 1398

\bibitem[{{Krabbe} {et~al.}(2004){Krabbe}, {Gasaway}, {Song}, {Iserlohe},
  {Weiss}, {Larkin}, {Barczys}, \& {Lafreniere}}]{Krabbe2004}
{Krabbe}, A., {Gasaway}, T., {Song}, I., {Iserlohe}, C., {Weiss}, J., {Larkin},
  J.~E., {Barczys}, M., \& {Lafreniere}, D. 2004, in Society of Photo-Optical
  Instrumentation Engineers (SPIE) Conference Series, Vol. 5492, Society of
  Photo-Optical Instrumentation Engineers (SPIE) Conference Series, ed.
  {A.~F.~M.~Moorwood \& M.~Iye}, 1403--1410

\bibitem[{{Larkin} {et~al.}(2006){Larkin}, {Barczys}, {Krabbe}, {Adkins},
  {Aliado}, {Amico}, {Brims}, {Campbell}, {Canfield}, {Gasaway}, {Honey},
  {Iserlohe}, {Johnson}, {Kress}, {Lafreniere}, {Magnone}, {Magnone},
  {McElwain}, {Moon}, {Quirrenbach}, {Skulason}, {Song}, {Spencer}, {Weiss}, \&
  {Wright}}]{Larkin2006}
{Larkin}, J., {Barczys}, M., {Krabbe}, A., {Adkins}, S., {Aliado}, T., {Amico},
  P., {Brims}, G., {Campbell}, R., {Canfield}, J., {Gasaway}, T., {Honey}, A.,
  {Iserlohe}, C., {Johnson}, C., {Kress}, E., {Lafreniere}, D., {Magnone}, K.,
  {Magnone}, N., {McElwain}, M., {Moon}, J., {Quirrenbach}, A., {Skulason}, G.,
  {Song}, I., {Spencer}, M., {Weiss}, J., \& {Wright}, S. 2006, New Astron.
  Rev., 50, 362

\bibitem[{{Macintosh} {et~al.}(2014){Macintosh}, {Graham}, {Ingraham},
  {Konopacky}, {Marois}, {Perrin}, {Poyneer}, {Bauman}, {Barman}, {Burrows},
  {Cardwell}, {Chilcote}, {De Rosa}, {Dillon}, {Doyon}, {Dunn}, {Erikson},
  {Fitzgerald}, {Gavel}, {Goodsell}, {Hartung}, {Hibon}, {Kalas}, {Larkin},
  {Maire}, {Marchis}, {Marley}, {McBride}, {Millar-Blanchaer}, {Morzinski},
  {Norton}, {Oppenheimer}, {Palmer}, {Patience}, {Pueyo}, {Rantakyro},
  {Sadakuni}, {Saddlemyer}, {Savransky}, {Serio}, {Soummer},
  {Sivaramakrishnan}, {Song}, {Thomas}, {Wallace}, {Wiktorowicz}, \&
  {Wolff}}]{Macintosh2014}
{Macintosh}, B., {Graham}, J.~R., {Ingraham}, P., {Konopacky}, Q., {Marois},
  C., {Perrin}, M., {Poyneer}, L., {Bauman}, B., {Barman}, T., {Burrows},
  A.~S., {Cardwell}, A., {Chilcote}, J., {De Rosa}, R.~J., {Dillon}, D.,
  {Doyon}, R., {Dunn}, J., {Erikson}, D., {Fitzgerald}, M.~P., {Gavel}, D.,
  {Goodsell}, S., {Hartung}, M., {Hibon}, P., {Kalas}, P., {Larkin}, J.,
  {Maire}, J., {Marchis}, F., {Marley}, M.~S., {McBride}, J.,
  {Millar-Blanchaer}, M., {Morzinski}, K., {Norton}, A., {Oppenheimer}, B.~R.,
  {Palmer}, D., {Patience}, J., {Pueyo}, L., {Rantakyro}, F., {Sadakuni}, N.,
  {Saddlemyer}, L., {Savransky}, D., {Serio}, A., {Soummer}, R.,
  {Sivaramakrishnan}, A., {Song}, I., {Thomas}, S., {Wallace}, J.~K.,
  {Wiktorowicz}, S., \& {Wolff}, S. 2014, Proceedings of the National Academy
  of Science, 111, 12661

\bibitem[{{Marley} {et~al.}(2012){Marley}, {Saumon}, {Cushing}, {Ackerman},
  {Fortney}, \& {Freedman}}]{Marley2012}
{Marley}, M.~S., {Saumon}, D., {Cushing}, M., {Ackerman}, A.~S., {Fortney},
  J.~J., \& {Freedman}, R. 2012, \apj, 754, 135

\bibitem[{{Marois} {et~al.}(2008){Marois}, {Macintosh}, {Barman}, {Zuckerman},
  {Song}, {Patience}, {Lafreni{\`e}re}, \& {Doyon}}]{Marois2008}
{Marois}, C., {Macintosh}, B., {Barman}, T., {Zuckerman}, B., {Song}, I.,
  {Patience}, J., {Lafreni{\`e}re}, D., \& {Doyon}, R. 2008, Science, 322, 1348

\bibitem[{{Marois} {et~al.}(2010){Marois}, {Zuckerman}, {Konopacky},
  {Macintosh}, \& {Barman}}]{Marois2010}
{Marois}, C., {Zuckerman}, B., {Konopacky}, Q.~M., {Macintosh}, B., \&
  {Barman}, T. 2010, \nat, 468, 1080

\bibitem[{{Metchev} {et~al.}(2009){Metchev}, {Marois}, \&
  {Zuckerman}}]{Metchev2009}
{Metchev}, S., {Marois}, C., \& {Zuckerman}, B. 2009, \apjl, 705, L204

\bibitem[{{Metchev} {et~al.}(2015){Metchev}, {Heinze}, {Apai}, {Flateau},
  {Radigan}, {Burgasser}, {Marley}, {Artigau}, {Plavchan}, \&
  {Goldman}}]{Metchev2015}
{Metchev}, S.~A., {Heinze}, A., {Apai}, D., {Flateau}, D., {Radigan}, J.,
  {Burgasser}, A., {Marley}, M.~S., {Artigau}, {\'E}., {Plavchan}, P., \&
  {Goldman}, B. 2015, \apj, 799, 154

\bibitem[{{{\"O}berg} {et~al.}(2011){{\"O}berg}, {Murray-Clay}, \&
  {Bergin}}]{Oberg2011}
{{\"O}berg}, K.~I., {Murray-Clay}, R., \& {Bergin}, E.~A. 2011, \apjl, 743, L16

\bibitem[{{Oppenheimer} {et~al.}(2013){Oppenheimer}, {Baranec}, {Beichman},
  {Brenner}, {Burruss}, {Cady}, {Crepp}, {Dekany}, {Fergus}, {Hale},
  {Hillenbrand}, {Hinkley}, {Hogg}, {King}, {Ligon}, {Lockhart}, {Nilsson},
  {Parry}, {Pueyo}, {Rice}, {Roberts}, {Roberts}, {Shao}, {Sivaramakrishnan},
  {Soummer}, {Truong}, {Vasisht}, {Veicht}, {Vescelus}, {Wallace}, {Zhai}, \&
  {Zimmerman}}]{Oppenheimer2013}
{Oppenheimer}, B.~R., {Baranec}, C., {Beichman}, C., {Brenner}, D., {Burruss},
  R., {Cady}, E., {Crepp}, J.~R., {Dekany}, R., {Fergus}, R., {Hale}, D.,
  {Hillenbrand}, L., {Hinkley}, S., {Hogg}, D.~W., {King}, D., {Ligon}, E.~R.,
  {Lockhart}, T., {Nilsson}, R., {Parry}, I.~R., {Pueyo}, L., {Rice}, E.,
  {Roberts}, J.~E., {Roberts}, Jr., L.~C., {Shao}, M., {Sivaramakrishnan}, A.,
  {Soummer}, R., {Truong}, T., {Vasisht}, G., {Veicht}, A., {Vescelus}, F.,
  {Wallace}, J.~K., {Zhai}, C., \& {Zimmerman}, N. 2013, \apj, 768, 24

\bibitem[{{Pueyo} {et~al.}(2014){Pueyo}, {Soummer}, {Hoffmann}, {Oppenheimer},
  {Graham}, {Zimmerman}, {Zhai}, {Wallace}, {Vescelus}, {Veicht}, {Vasisht},
  {Truong}, {Sivaramakrishnan}, {Shao}, {Roberts}, {Roberts}, {Rice}, {Parry},
  {Nilsson}, {Luszcz-Cook}, {Lockhart}, {Ligon}, {King}, {Hinkley},
  {Hillenbrand}, {Hale}, {Dekany}, {Crepp}, {Cady}, {Burruss}, {Brenner},
  {Beichman}, \& {Baranec}}]{Pueyo2014}
{Pueyo}, L., {Soummer}, R., {Hoffmann}, J., {Oppenheimer}, R., {Graham}, J.~R.,
  {Zimmerman}, N., {Zhai}, C., {Wallace}, J.~K., {Vescelus}, F., {Veicht}, A.,
  {Vasisht}, G., {Truong}, T., {Sivaramakrishnan}, A., {Shao}, M., {Roberts},
  Jr., L.~C., {Roberts}, J.~E., {Rice}, E., {Parry}, I.~R., {Nilsson}, R.,
  {Luszcz-Cook}, S., {Lockhart}, T., {Ligon}, E.~R., {King}, D., {Hinkley}, S.,
  {Hillenbrand}, L., {Hale}, D., {Dekany}, R., {Crepp}, J.~R., {Cady}, E.,
  {Burruss}, R., {Brenner}, D., {Beichman}, C., \& {Baranec}, C. 2014, ArXiv
  e-prints

\bibitem[{{Sadakane}(2006)}]{Sadakane2006}
{Sadakane}, K. 2006, \pasj, 58, 1023

\bibitem[{{Skemer} {et~al.}(2012){Skemer}, {Hinz}, {Esposito}, {Burrows},
  {Leisenring}, {Skrutskie}, {Desidera}, {Mesa}, {Arcidiacono}, {Mannucci},
  {Rodigas}, {Close}, {McCarthy}, {Kulesa}, {Agapito}, {Apai}, {Argomedo},
  {Bailey}, {Boutsia}, {Briguglio}, {Brusa}, {Busoni}, {Claudi}, {Eisner},
  {Fini}, {Follette}, {Garnavich}, {Gratton}, {Guerra}, {Hill}, {Hoffmann},
  {Jones}, {Krejny}, {Males}, {Masciadri}, {Meyer}, {Miller}, {Morzinski},
  {Nelson}, {Pinna}, {Puglisi}, {Quanz}, {Quiros-Pacheco}, {Riccardi},
  {Stefanini}, {Vaitheeswaran}, {Wilson}, \& {Xompero}}]{Skemer2012}
{Skemer}, A.~J., {Hinz}, P.~M., {Esposito}, S., {Burrows}, A., {Leisenring},
  J., {Skrutskie}, M., {Desidera}, S., {Mesa}, D., {Arcidiacono}, C.,
  {Mannucci}, F., {Rodigas}, T.~J., {Close}, L., {McCarthy}, D., {Kulesa}, C.,
  {Agapito}, G., {Apai}, D., {Argomedo}, J., {Bailey}, V., {Boutsia}, K.,
  {Briguglio}, R., {Brusa}, G., {Busoni}, L., {Claudi}, R., {Eisner}, J.,
  {Fini}, L., {Follette}, K.~B., {Garnavich}, P., {Gratton}, R., {Guerra},
  J.~C., {Hill}, J.~M., {Hoffmann}, W.~F., {Jones}, T., {Krejny}, M., {Males},
  J., {Masciadri}, E., {Meyer}, M.~R., {Miller}, D.~L., {Morzinski}, K.,
  {Nelson}, M., {Pinna}, E., {Puglisi}, A., {Quanz}, S.~P., {Quiros-Pacheco},
  F., {Riccardi}, A., {Stefanini}, P., {Vaitheeswaran}, V., {Wilson}, J.~C., \&
  {Xompero}, M. 2012, \apj, 753, 14

\bibitem[{{Smith}(1998)}]{Smith1998}
{Smith}, M.~D. 1998, Icarus, 132, 176

\bibitem[{{Visscher} \& {Fegley}(2005)}]{Visscher2005}
{Visscher}, C. \& {Fegley}, Jr., B. 2005, \apj, 623, 1221

\bibitem[{{Visscher} {et~al.}(2010){Visscher}, {Moses}, \&
  {Saslow}}]{Visscher2010}
{Visscher}, C., {Moses}, J.~I., \& {Saslow}, S.~A. 2010, Icarus, 209, 602

\bibitem[{{Warmbier} {et~al.}(2009){Warmbier}, {Schneider}, {Sharma}, {Braams},
  {Bowman}, \& {Hauschildt}}]{Warmbier2009}
{Warmbier}, R., {Schneider}, R., {Sharma}, A.~R., {Braams}, B.~J., {Bowman},
  J.~M., \& {Hauschildt}, P.~H. 2009, \aap, 495, 655

\bibitem[{{Wong} {et~al.}(2004){Wong}, {Mahaffy}, {Atreya}, {Niemann}, \&
  {Owen}}]{Wong2004}
{Wong}, M.~H., {Mahaffy}, P.~R., {Atreya}, S.~K., {Niemann}, H.~B., \& {Owen},
  T.~C. 2004, \icarus, 171, 153

\bibitem[{{Yurchenko} \& {Tennyson}(2014)}]{Yurchenko2014}
{Yurchenko}, S.~N. \& {Tennyson}, J. 2014, \mnras, 440, 1649

\bibitem[{{Zahnle} \& {Marley}(2014)}]{Zahnle2014}
{Zahnle}, K.~J. \& {Marley}, M.~S. 2014, \apj, 797, 41

\end{thebibliography}
\end{document}